%% file: fusion.tex
\documentclass[prodmode,acmtoms]{acmsmall}

\input{mydef}

\begin{document}

\markboth{P. D'Alberto}{OpenCL+APU+GPU+MM}

\title{A Heterogeneous Accelerated Matrix Multiplication: OpenCL + APU + GPU+ Fast
  Matrix Multiply}

\author{PAOLO D'ALBERTO
  \affil{FastMMW, CA, USA}
}

\begin{abstract}
  As users and developers, we are witnessing the opening of a new
  computing scenario: the introduction of hybrid processors into a
  single die, such as an accelerated processing unit (APU) processor,
  and the plug-and-play of additional graphics processing units (GPUs)
  onto a single motherboard. These APU processors provide multiple
  symmetric cores with their memory hierarchies and an integrated
  GPU. Moreover, these processors are designed to work with external
  GPUs that can push the peak performance towards the TeraFLOPS boundary.
  
  We present a case study for the development of dense Matrix
  Multiplication (MM) codes for matrix sizes up to 19K$\times$19K, thus
  using all of the above computational engines, and an achievable peak
  performance of 200 GFLOPS for, literally, a made-at-home built. We
  present the results of our experience, the quirks, the pitfalls, the
  achieved performance, and the achievable peak performance.
\end{abstract}




\acmformat{P. D'Alberto. OpenCL+APU+GPU+FastMM}

\begin{bottomstuff}
Author's addresses: P. D'Alberto paolo@FastMMW.com
\end{bottomstuff}

\maketitle

\section{Introduction}
 
As users and consumers, we are accustomed to having multiple cores in
a single processor and we are enjoying the many advantages.  Nowadays,
processors with two or more cores are common in notebooks, tablets and
smart phones, delivering additional performance.  Desktops may have
4-8 core processors, servers usually  have multi-core processors
as well.  Also, as occasional gamers, either through casual games
played in a browser or multiplayer games played on a console or PC,
graphics processing units provide those realisitic effects we are used
to and take for granted.

As developers and algorithm designers, we are experiencing a kind of
Renaissance because we are stimulated to design algorithms to exploit
these new computational engines for both new and old applications. A
Renaissance indeed, because super computing is not anymore at the
fingertips of only a small elite but it is practically for
everyone. Think, anyone capable to use a screw driver could build a
desktop capable to deliver one and more TeraFLOPS peak performance
with a few GPUs in it \cite{VetterGDSLMMRRSY2011}; paraphrasing Cray's
saying: we have a few oxes pulled by hundred of chickens. The last
attempt to do such a popularization of supercomputing was by the Cell
processor and the PS3 game console (which became impossible for future
systems as SONY removed support for the LINUX operating system)
achieving the same performance by the same flop per dollar ratio that
we shall present in this work.

In this work, we turn our attention to heterogeneous systems and in
particular to single board systems with hybrid processors, that is
with symmetric cores and a GPU, and additional external GPUs; we call
these {\em computational engines}; each computation engine will have
very different performances and will fit very different computational
needs. Here, we can easily add or substitute computational engines in
the system: for example, we can change a GPU by a snap (or two) and we
want software to change the work load accordingly even at run
time. Note, in this type of systems, the GPU is one component. In
particular, the external GPUs can be omitted altogether and still have
GPU capabilities. Moreover, as the technology improves, we may easily
pluck out the APU processor for a new version, with larger GPU within
or more cores. This upgrade of the system is more in line with small
budgets planning, where only a part of the machine is upgraded, not
decommissioned, and the rest is left unchanged. The ability to write
code that, in principle, adapts to the different configurations with
little or no modifications will make these systems even more
appealing: simplifying costly software maintenance.

We neither measure nor present in-GPU timing (AMD OpenCL package
provides examples of how to measure the internal computation time only
but we wanted to measure the so called wall-clock as well). We take
the point of view that GPUs and APUs are accelerating devices, thus we
should present the overall acceleration in combination with classic
computation (non-accelerated or CPU-only) so to appreciate the organic
performance. Of course, the performance will be less jaw dropping, it
will be sober and reasonable, nonetheless impressive.  After all, we
are interested in those types of computations where the transfer of
data and its execution time (of the transfer) are integral parts of
the computation. To be fair, we measure performance for problem sizes
that are very large and they will not fit in any computation-engine
internal memory.

We will take an agnostic view of the GPUs and the code for them. In
fact, we are going to use OpenCL to abstract the system resources and
we will use the OpenCL interface to guide the computations.  Also, we
are going to take the MM Kernels provided within the OpenCL samples as
{\em they are}. What we are after is the ability to determine the
capacity storage of the GPUs or internal memory. Thus, we are
interested in the workload capability of the GPUs and we will reuse
the code available.  We shall go into the details in in Section
\ref{sec:recursive-description}.

We are going to use a different attitude about the code for the
CPUs. We will reuse the code provided by ATLAS and GotoBLAS. That is,
we are going to use the best known codes for multi-core systems.  We
will deploy with our codes the SGEMM'S from ATLAS because of a conflict
with thread allocation using GotoBLAS. However, we will provide the
performance for both so that to appreciate the hardware-accelerators
effects.

We choose to present performance for the Matrix Multiplication kernel
because: First, it is a well known kernel; second, there are close to
optimal codes for both GPUs and CPUs; third, we are interested in the
interrelation among CPUs and GPUs, which is a relatively new problem;
and fourth, we are interested to investigate how close we can get to
the peak performance.

The challenges to solve are not new and they are not trivial
either. We shall show a natural and simple approach to take advantage
of the diversity of the computational engines and we shall show that
all engines are useful in different ways: First, CPUs will provide the
best solution for relatively small problems; second, all GPUs should
be used for the solution of intermediate and large problems; finally,
CPUs will support coordination and data-layout transformations
necessary for the handling of very large problems.

We organize our work as it follows. In Section \ref{sec:related-work},
we shall try our best in providing a survey about the related work. In
Section \ref{sec:top-down}, we shall introduce our contribution and
system in a top-down fashion: in Section
\ref{sec:recursive-description}, we shall present the recursive
algorithm that will break down the computation in smaller ones to be
solved by the computational engines; in Section
\ref{sec:leaf-computation}, we shall provide the details how we
combine the power of the different engines; in Section
\ref{sec:opencl-configuration}, we shall describe how we use OpenCL to
abstract the computational engines, and in Section \ref{sec:Hardware}
we shall provide details about the hardware we deployed. In Section
\ref{sec:experimental-results}, we shall present our experimental
results: as peak performance, in Section \ref{sec:peak-performance},
and as achievable performance, in Section
\ref{sec:achievable-performance}.  We conclude with our
acknowledgments.

A note: In this work, we will not discuss nor present any numerical
analysis such as maximum error, maximum relative error.

\section{Related Work}
\label{sec:related-work}
We can divide this section into several parts: For example, about the
Matrix Multiplication and its applications, about implementations of
MM for multi-core multi-processors, about implementation for GPUs,
Software/Hardware hybrid implementations where desktop solutions are
combined with low power field programmable gates FPGAs solutions.  In
fact, MM is so ubiquitous in science that it is used very often as
kernel, as a basic operation, and also as a benchmark for new systems,
for new architectures. This exposure of MM in different fields and the
simplicity how MM can be presented make MM like a common language and
often it is taken for granted; at the same time, it is also like a
secret hand-shake for researcher communities, among who a very few
researchers have mastered it really.

Matrix Multiplication is considered such an old-school problem but it
still attracts a large volume of research. We are all familiar with
the algorithm of complexity $O(N^3)$, which is the standard
implementation in the NETLIB BLAS. In turn, BLAS 3, the set of
matrix-matrix operation can be reduced to matrix multiplication
\cite{KagstromLVL981,BlackfordDDDHHHKLPPRW2002}. Optimized BLAS are
extremely useful and ubiquitous in scientific and statistical software
packages, often we may used them without know it. In this work, we
work with ATLAS \cite{WhaleyD98} and with GotoBLAS \cite{GotoG2006}.  We
are interested in the so called Fast MM algorithms:
Vassilevska-Williams \cite{Vassilevska2011} Coppersmith-Winograd's
\cite{CoppersmithW87}, Pan's \cite{Pan1978}, Strassen's
\cite{Strassen69}, and our recent implementations for SMP machines
\cite{Dalberto:2011:EPM}. Fast MM are practical and stable
\cite{DemmelDHK2006,DemmelH92}. The connection between MM and other
applications can be surprising: for example in a semi-ring (where the
$+$ operation may not have inverse) the All-Pair shortest-path and the
classic MM are computationally equivalent and they have the same
solution \cite{DAlbertoNRKleene2007,Warshall62,Floyd62}. The
connection was not lost in the implementation for GPUs \cite{Buluc08}.
 
MM has been used a benchmark or as motivational example for compiler
optimizations such as tiling, parallelism by threads manually or
automatically by OpenMP \cite{ChandraDKMMM2000} or Cilk
\cite{FrigoHLL09}. Other optimizations in combination with the above
is matrix layout optimizations \cite{ChatterjeeLPT2002}, which we
could take advantage in this work as well.

The parallelism of MM is a central subject of this work. For Fast
MM, the authors have a recent contribution to exploit the full speed
up in symmetric multi-core processors \cite{Dalberto:2011:EPM}. In that
work, symmetry of the architecture and of the algorithm is fundamental
for achieving the best performance. Here, in contrast, asymmetric
computational engines are part of the architecture. The software must
be aware and adapt. 

Currently, GPUs are having more and more traction in the scientific
computing as flexible means to compute complex algorithms and,
especially, as computational engine with jaw-dropping performance
reaching easily 500GFLOPS. The literature is rich of fast GEMM
implementations of MM
\cite{Tan:2011:FID:2063384.2063431,Li:2009:NAG:1561015.1560858,Volkov:2008:BGT:1413370.1413402},
Fast MM \cite{LiRS2011}, and fast and accurate
\cite{Badin:2011:IAM:1971897.1971901}.

What actually attracted us to this subject ---i.e., MM for
heterogeneous computing--- has been the arrival on a new architecture
such as the APU \cite{APUwhitepaper2010} and the OpenCL
\cite{GasterHKMS2011} as a programming environment/API. Using OpenCL,
AMD OpenCL examples, and a little practice, we could run code for both
the CPUs and the GPUs, without knowing the inner workings of
neither. Considering that the high performance codes for the CPUs took
a decade to reach the level they are now, the ability to write code
for even more complex system is quite something.

From the exchange of emails with other researchers who have asked for
our Fast MM codes, we have noticed that such an ease to code can tempt
developer to use a single code for all devices. This is a cheap
solution, the code is easy, there is no maintenance but the
performance and efficiency will be poor defying the purpose of these
beautiful machines.

\section{Top-Down Matrix Multiplication}
\label{sec:top-down}

We opt to present our system in a top-down fashion. We start by
presenting the classic recursive algorithm for the computation of
Matrix Multiplication (Section \ref{sec:recursive-description}). The
recursive algorithm, when reaching an appropriate problem size, will
yield control to a leaf computation (Section
\ref{sec:leaf-computation}). The leaf computation can be a CPU only
computation, GPUs only computation, and GPUs and CPUs. The leaf
computation is based on an abstraction of the computational engines as
we present in Section \ref{sec:opencl-configuration}. We describe the
hardware of our system, the possible configurations and we show a
picture of the build in Section \ref{sec:Hardware}.

\subsection{Recursive Description}
\label{sec:recursive-description}

Our goal is to compute matrix multiplication for any problem size with
the help of different computational engines. In this work, we use a
recursive algorithm that is designed to divide the problem in similar
sub-problems using a recursive formulation.  We actually have two
recursive algorithm explicitly computing the single MM and the
multiply-add matrix computation, see Table \ref{tab:rmul}. 

Note that the algorithm stems from the observation that any matrix
$\Vc{D} \in \R^{m\times n}$ can be always divided into four quadrants:
\begin{equation}
  \label{eq:matrix-division}
  \Vc{D} =
  \begin{bmatrix}
    \Q{D}{0} & \Q{D}{1} \\
    \Q{D}{2} & \Q{D}{3} \\
  \end{bmatrix}.
\end{equation}
Here, we divide the matrix so that $\Vc{D}_0 \in
\R^{\lceil\frac{m}{2}\rceil\times\lceil\frac{n}{2}\rceil}$ and
$\Vc{D}_3 \in
\R^{\lfloor\frac{m}{2}\rfloor\times\lfloor\frac{n}{2}\rfloor}$.
\begin{table}
 \tbl{Matrix Multiplication $\Vc{C} = \Vc{A}\Vc{B}$ descriptions. \label{tab:rmul}} {%
   \label{tab:lmul}
   \begin{tabular}{|l|l|}
     \hline \hline $\Vc{C} = \Vc{A}\Vc{B}$ & $\Vc{C}{+=}\Vc{A}\Vc{B}$  \\ \hline \hline
     if small then Leaf($\Vc{C},\Vc{A},\Vc{B}$) & if small then LeafAdd($\Vc{C},\Vc{A},\Vc{B}$) \\
     else & else \\
     \hspace{0.5cm} $\Q{C}{0}=   \Q{A}{0}\Q{B}{0}$ &\hspace{0.5cm} $\Q{C}{0}{+=}\Q{A}{0}\Q{B}{0}$  \\
     \hspace{0.5cm} $\Q{C}{1}=   \Q{A}{0}\Q{B}{1}$ &\hspace{0.5cm} $\Q{C}{1}{+=}\Q{A}{0}\Q{B}{1}$  \\
     \hspace{0.5cm} $\Q{C}{2}=   \Q{A}{2}\Q{B}{0}$ &\hspace{0.5cm} $\Q{C}{2}{+=}\Q{A}{2}\Q{B}{0}$  \\
     \hspace{0.5cm} $\Q{C}{3}=   \Q{A}{2}\Q{B}{1}$ &\hspace{0.5cm} $\Q{C}{3}{+=}\Q{A}{2}\Q{B}{1}$  \\
     \hspace{0.5cm} $\Q{C}{0}{+=}\Q{A}{1}\Q{B}{2}$ &\hspace{0.5cm} $\Q{C}{0}{+=}\Q{A}{1}\Q{B}{2}$  \\
     \hspace{0.5cm} $\Q{C}{1}{+=}\Q{A}{1}\Q{B}{3}$ &\hspace{0.5cm} $\Q{C}{1}{+=}\Q{A}{1}\Q{B}{3}$  \\
     \hspace{0.5cm} $\Q{C}{2}{+=}\Q{A}{3}\Q{B}{2}$ &\hspace{0.5cm} $\Q{C}{2}{+=}\Q{A}{3}\Q{B}{2}$  \\
     \hspace{0.5cm} $\Q{C}{3}{+=}\Q{A}{3}\Q{B}{3}$ &\hspace{0.5cm} $\Q{C}{3}{+=}\Q{A}{3}\Q{B}{3}$  \\ \hline
 \end{tabular}
 }
\end{table}
In this work, we shall present results for square matrices, however
the recursive algorithm is oblivious of the shape of the
matrices. Furthermore, the division of matrices into balanced
sub-matrices is the foundation of fast algorithm and thus we could
always use a fast algorithm presented in previous work without any
modification of the leaf computation. But this is beyond the scope of
this investigation.

The goal of a balanced recursive algorithm is simplicity and recursive
tiling. In contrast, tiling of the classic MM divides the problem into
smaller sub-problems and mostly of fixed size. Classic tiling could
provide better performance for a given architecture but less
flexibility. At this level of the computation, we rather have the
latter and compromise a little with the former.

In Table \ref{tab:rmul}, we omitted the details of when the problem
size is {\em small}. In naive terms, we would like to yield to the
leaf computation when either the CPUs or the GPUs can handle the
problem at hand directly. In this work, the recursion stops when the
operands matrix size is smaller than $6016\times6016$, this is also
called {\bf recursion point}. We shall dwell into the details in the
experimental section, when the architecture will be
clear. Intuitively, the recursion is chosen so that both the internal
GPU and the external GPU can provide almost peak performance.

\subsection{Leaf Computation}
\label{sec:leaf-computation}

We turn our attention to the leaf computation that performs the MM
$\Vc{C} = \Vc{A}\Vc{B}$. The leaf computation is simple to describe:

\begin{table}[htb]
  \tbl{Leaf(C,A,B) \label{tab:leaf}} {%
    \begin{tabular}{l}
      \hline \hline Leaf$(\Vc{C},\Vc{A},\Vc{B})$ \\ \hline \hline
      if size  $\geq K_1$ \\
      \hspace{0.5cm}  rGPUs(C,A,B)\\ 
      if size  $\geq K_0$  and size $< K_1$ \\
      \hspace{0.5cm}  GPU(C,A,B)\\ 
      otherwise \\
      \hspace{0.5cm}  SGEMM(C,A,B)\\
      \hline 
    \end{tabular}
  }
\end{table}

Let us recall that we are working with a system composed of an APU and
an external GPU. That is, we have two GPUs, one is internal to the APU
and the other is external.

If the problem size is larger than a critical point, we will use
either a single or multiple GPUs to solve the problem; otherwise, we
call SGEMM (from any high performance BLAS 3 library).

Let us address the small problem first. Experimentally and for this
architecture, if the matrices are smaller than $400 \times 400$ (i.e.,
$K_0=400$) we are better off using SGEMM: we took in consideration
both ATLAS and GotoBLAS. We eventually decided to use ATLAS because
GotoBLAS affects the thread allocation in OpenCL adversely by
serializing the GPU computations.  However, we shall show that
GotoBLAS SGEMM standalone is faster than the ATLAS counterpart. In the
experimental section, we shall provide more details.

For problem sizes larger than $K_0=400$ and smaller than $K_1 \sim
3000$ we will use a single GPU, the external one. In the following,
Section \ref{sec:opencl-configuration}, we will provide the details
the GPU MM kernel. The choice of the breaking point $3000$ is small
with respect to the capacity of the external GPU, which can solve MM
with matrices up to $4300\times 4300$. The size $3000$ is the
break-even point when both GPUs should work collaboratively and also
the problem size that the internal GPU can solve directly. In our
system, the GPU crossfire is activated, thus boosting the throughput
and thus the parallelism between the GPUs.

Now, let us address the computation using GPUs, which is at the center
of our work. Once again the idea is simple. Consider the problem
$\Vc{C} = \Vc{A}\Vc{B}$, we can split the matrices as follows.
\begin{equation}
  \begin{bmatrix}
    \Q{C}{0} & \Q{C}{1} \\
    \Q{C}{2} & \Q{C}{3} \\
  \end{bmatrix} = 
  \begin{bmatrix}
    \Q{A}{0} & \Q{A}{1} \\
    \Q{A}{2} & \Q{A}{3} \\
  \end{bmatrix}*
  \begin{bmatrix}
    \Q{B}{0} & \Q{B}{1} \\
    \Q{B}{2} & \Q{B}{3} \\
  \end{bmatrix}.
  \label{eq:matrix-division-computation}
\end{equation}
and thus we can allocate to one GPU the following computation:
\begin{equation}
  \Q{C}{0} = \Q{A}{0}*\Q{B}{0};\hspace{0.2cm}
  \Q{C}{0} += \Q{A}{1}*\Q{B}{2};\hspace{0.2cm}
  \Q{C}{2} = \Q{A}{2}*\Q{B}{0};\hspace{0.2cm}
  \Q{C}{2} += \Q{A}{3}*\Q{B}{2}
  \label{eq:matrix-division-gpu0}
\end{equation}
and to the other the smaller computation:
\begin{equation}
  \Q{C}{1} = \Q{A}{0}*\Q{B}{1};\hspace{0.2cm}
  \Q{C}{1} += \Q{A}{1}*\Q{B}{3};\hspace{0.2cm}
  \Q{C}{3} = \Q{A}{2}*\Q{B}{1};\hspace{0.2cm}
  \Q{C}{3} += \Q{A}{3}*\Q{B}{3}.
  \label{eq:matrix-division-gpu1}
\end{equation}

The matrices do not need to be square. Nonetheless, the computation is
balanced, in the sense that $\Q{C}{0} = \Q{A}{0}*\Q{B}{0}$ computes
just $N^2$ operations more than $\Q{C}{3} += \Q{A}{3}*\Q{B}{3}$, where
$N\times N$ is the matrix size of $\Q{C}{0}$. This difference is in
the matrix-vector computations needed to compute the borders of
$\Q{C}{0}$, which account for $2N-1$ extra elements.

In general, this does not need to be: one GPU could work on a problem
much larger than the other. We tested such an unbalance division but,
for our system, it did not provide any performance advantage and thus
we do not discuss it any further.

A simple optimization, for which we will present results in a future
work, is the change of layout of the operands so that the sub matrices
---i.e., $\Q{C}{0},\Q{C}{1},\Q{C}{2}$, and $\Q{C}{3}$--- are
continuous in memory. The advantage is two fold: the change of layout
is used by both GPU computations, thus we can save communications, and
this will speed up the communication between memory and the GPU
internal memory.

\subsection{OpenCL Configuration}
\label{sec:opencl-configuration}
We abstract our system by using platforms, queues, and devices. A {\bf
  platform} is composed by devices: CPUs and GPUs. A platform can have
multiple {\bf devices}: In this work we work with an internal CPU
device, an internal GPU device (we shall use the term $GPU_1$), and an
external GPU device ($GPU_0$). In particular, we use OpenCL to
abstract only GPU devices. Any device is identified by an unique
integer and the basic information about the device can be queried
using this unique identifier. We associate to every GPU device a
Matrix Multiplication {\bf queue} structure.

A MM queue will collect basic information about the device such as the
size of the internal memory, if it has any.  The MM queue will have
the function of an OpenCL queue: Memory context, memory buffers,
programs or kernels, and command queue.

We built a system that has three devices (CPU, $GPU_1$ and
$GPU_0$). We consider the two GPUs like a priority queue: where we
serve first and with larger problem the $GPU_0$ and then $GPU_1$. We
query the information about the devices and in particular we determine
the internal data buffer in such a way we estimate the problem size
that a device will be able to solve. The internal $GPU_1$ can store
three matrices of size $3008\times 3008$, thus a square problem size
of $N=3008$. The external $GPU_0$ can solve a larger problem: that is,
$N=4305$. Such a capacity of the GPUs is fundamental for the division
strategy and it is determined and exploited at run time.

To initialize a MM queue, we create a context and a command queue
first. A {\bf context} is an abstract object that manages the
interaction between the host (program) and the devices such as memory
objects in a device and kernel programs created for a device. A {\bf
  command queue} is the main mechanism to communicate, to start a
computation, and to control a device. Then, we create the buffers and
the codes.  We create three {\bf buffers}, which are contiguous memory
that the device uses as transient memory to read the matrix operands
and to write the matrix result. We compile OpenCL programs for MM that
are available through the AMD OpenCL distribution. We modified the
programs a little to adapt to a few new requirements, but the
modifications are minor.
 
Once the MM queue is initialized, we are ready to execute commands; for
example, the matrix computation $\Vc{C}+= \Vc{A}\Vc{B}$ is performed
by three basic MM queue routines in order:
\begin{enumerate}
\item Move the operands $\Vc{A}$ and $\Vc{B}$ into the input buffers
  and wait for the communication completion.
\item Execute the basic MM kernel, $\Vc{C} = \Vc{A}\Vc{B}$, which is
  like an external function call.
\item Move the output into a local memory space and add $\Vc{C}$.
\end{enumerate}

We believe that the communication operations (moving the input
matrices and output matrix) are intuitive to grasp without too many
details, especially if the matrix operands are contiguous in
memory. We believe that the kernel execution is less intuitive and it
deserves more details.

First, let us recall that we are working with GPU engines and thus the
computation should be designed for this graphic unit. As intuitive
description, we can divide the computation into three parts: the
splitting of the computation into threads, the instantiation of the
parameters, and then the actual queuing of the kernel for
execution. Of course, we will wait for the completion of the execution
before to return. This is identical, at least in principle, to a
function call. 

The main difference with a function call is the initial division of
the original computation into threads. This division is only nominal
in the sense that is not apparent at code level but the GPUs internal
mechanism will carry it on: Consider the result matrix $\Vc{C}$, if we
divide the matrix into four quadrants as previously, we can divide the
computation into four independent computations or threads, each thread
will compute a blocked row-by-column matrix multiplications such as
\begin{equation}
\Q{C}{0} = \Q{A}{0}*\Q{B}{0}+\Q{A}{1}*\Q{B}{2}.
\end{equation}

The number of threads, that is, how we divide the matrix $\Vc{C}$,
depends on the type of GPUs basically. Each computation is often
implemented by the classic MM row-by-column operation. In principle,
each thread could be computing a single point of $\Vc{C}$ like this:
\begin{equation}
c_{i,j} = \sum_{k=1}^N a_{i,k}b_{k,j} 
\end{equation}
In this work, the kernel computes a $32\times 32$ tile of
$\Vc{C}$. This division process is natural in the field of loop
parallelization: we parallelize the outer loop of the classic MM
(three for-loops), which is common using OpenMP pragmas.  This is not
necessarily the best strategy and a blocked version could provide
better raw FLOPS performance and also achieving smaller numerical
error.

\subsection{Hardware}
\label{sec:Hardware}

The system built has the following specifications
\begin{itemize}
\item 16GB Memory  \verb{4Gx4|CORSAIR CMZ8GX3M2A1866C9B2Z{
\item Motherboard  \verb{ASUS|F1A75-M PRO R1{
\item Processor \verb2A8-38502
  \begin{itemize}
  \item 4 CPUs
  \item AMD Radeon HD 6550
  \item Off-market CPU cooler Hyper 212 plus
  \end{itemize}
\item External GPU, Diamond Radeon HD 6570
\end{itemize}

In Figure \ref{fig:built}, we show literally a snapshot of the built
system. 
\singlefigure{0.8}{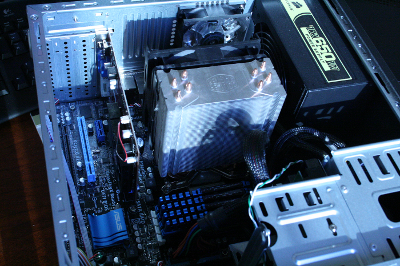}{Board snapshot}{fig:built}

Through the BIOS of the ASUS motherboard, we can set the default base
clock and thus configure the system. We tried a few, in Table
\ref{tab:configurations}, we show the ones stable so that we could run
experiments.

\begin{table}
  \tbl{Configurations \label{tab:configurations}} {%
    \begin{tabular}{|r|r|r|r|r|r|r|}
      \hline \hline 
      Base Clock mult.& Default & 90   & 100  & 112  & 114  & 115 \\ \hline \hline
      APU MHz (peak)  & 2900    & 2610 & 2900 & 3248 & 3306 & 3335 \\
      Memory MHz      & 1333    & 1440 & 1600 & 1792 & 1824 & 1840 \\ \hline
      \end{tabular}
  }
\end{table}

In the following of the paper, we will use the base clock multiplier
(i.e., Default, 90, etc) so that to identify the system and present
performance. We wrote all the codes independently of the hardware
configurations. In practice, we wrote the codes and tune them in the
default configuration.  

We installed Ubuntu 10.04 Natty, we then installed ATI Catalyst 11.8
and AMD OpenCL pack. The code used in this work is a variation of the
already available in the samples distribution. We will provide our
codes if requested.

\section{Experimental Results}
\label{sec:experimental-results}
For presentation purposes, we split the section into three
subsections.  We start with the software only performance; in Section
\ref{sec:MMresults}, we present the performance of GEMM by ATLAS and
GotoBLAS. Then, we show the peak performance of the system using only
GPUs, in Section \ref{sec:peak-performance}, and what we can achieve
when all data transfers are considered, see Section
\ref{sec:achievable-performance}.  Notice that when we talk about peak
performance, we do not consider in-GPU timing and we do not consider
hypothetical performance by resources counting and throughput only.

We measure performance as GLOPS (giga floating point operation per
second). We measure the ratio of the number of operation divided by
the wall-clock execution time of the MM. The number of operations is
$2N^3$ where $N$ is the problem size. We consider square problems for
presentation purpose, that is, convenience.

\subsection{Peak Performance:  CPUs only}
\label{sec:MMresults}

The CPU device in the APU provides a four-core system that can be used
to run efficient implementations of SGEMM. In this section, we present
the performance for SGEMM from the ATLAS and from the GotoBLAS
library. Also, we present the performance for the Winograd's MM as
implemented by the same authors in \cite{Dalberto:2011:EPM} and we shall use the symbol
SW. We will show that the same algorithm called in the OpenCL
environment will have different performance.  The performance
presented in this section is measured independently of the OpenCL and
its framework.

\doublefigure{0.48}{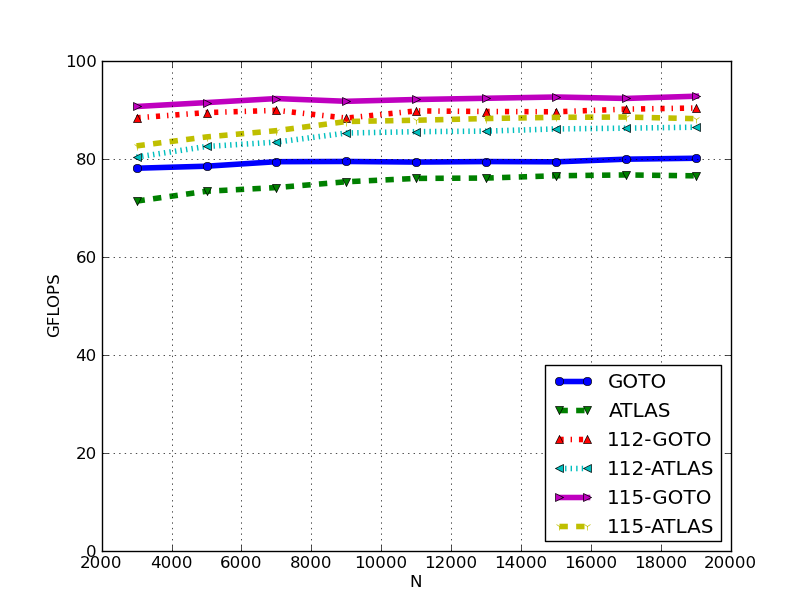}{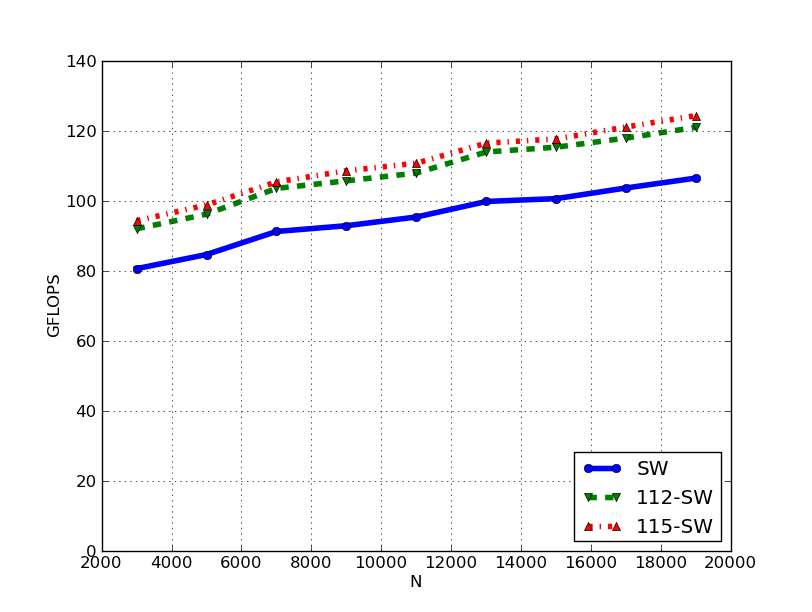}{Parallel CPUs: GotoBLAS's, ATLAS's SGEMM and
  Goto-based Winograd's MM performance}{fig:cpus-performance}

In Figure \ref{fig:cpus-performance}, we can see that the GotoBLAS
SGEMM is faster than ATLAS's SGEMM. For GotoBLAS, we generate code
for the Shanghai architecture (i.e., \verb1GotoBLAS21) because the APU
processor is not recognized in current installation process. ATLAS's
is self tuned and it provides very good performance. ATLAS's SGEMM is
about 5\% slower for larger problems and about 10\% slower for smaller
ones. Our implementation of Winograd implementation is based on Goto's
SGEMM so that to show what could it be the performance by using
Algorithm acceleration only.

We can see that SGEMM implemented with the best code for this APU
cores run at about 90 GFLOPS. We shall show this performance is about
20 GFLOPS slower that the MM using the internal GPU alone, 30 GFLOPS
slower than using the external GPU, and 60 GFLOPS slower than using
all. Notice also that we can achieve about 120 GFLOPS using Winograd's
implementation: making it as fast as the internal GPU, which is
very competitive.

\subsection{Peak Performance: GPUs only}
\label{sec:peak-performance}

In this section, we address the peak performance that we can achieve
using only the GPUs. In particular, we have the matrix operands stored
continuously in memory, thus requiring little or no
pre-computation. In this way, we can measure the peak performance of
the GPUs when the data reside in memory (off the GPUs local
storage). In Figure \ref{fig:gpus-performance-0}, we show the
performance we can achieve for every GPU separately with different
configurations.

\doublefigure{0.48}{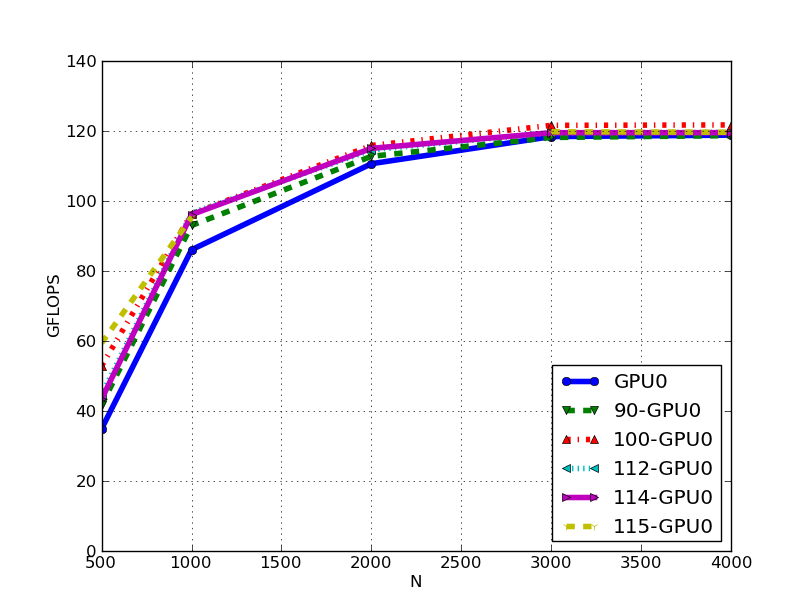}{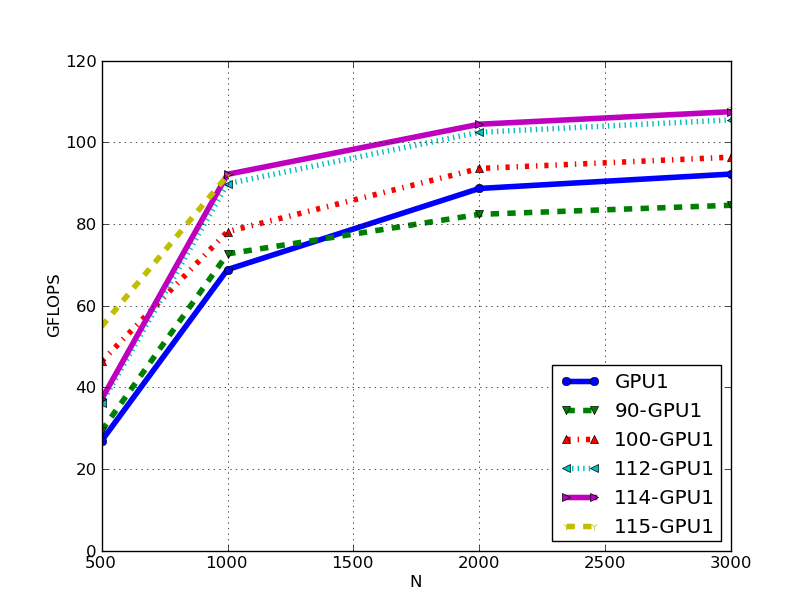}{Each GPU performance
  respectively}{fig:gpus-performance-0}

We recall the notation used: The $GPU_0$ is the external GPU
(connected through the PCI) and it can solve directly problems of size
up to $4300\times 4300$. The $GPU_1$ is the internal GPU and it can
compute directly problems of sizes up to $3008\times 3008$. 

We notice that $GPU_0$ can achieve up to 120GFLOPS peak performance
independently of the system configuration. However, for smaller
problems a faster memory allows better performance. In contrast, the
$GPU_1$ improves consistently as the configuration gets faster. There
is about 10--30 GFLOPS performance difference between the two GPUs as
a function of the configuration.

\singlefigure{0.7}{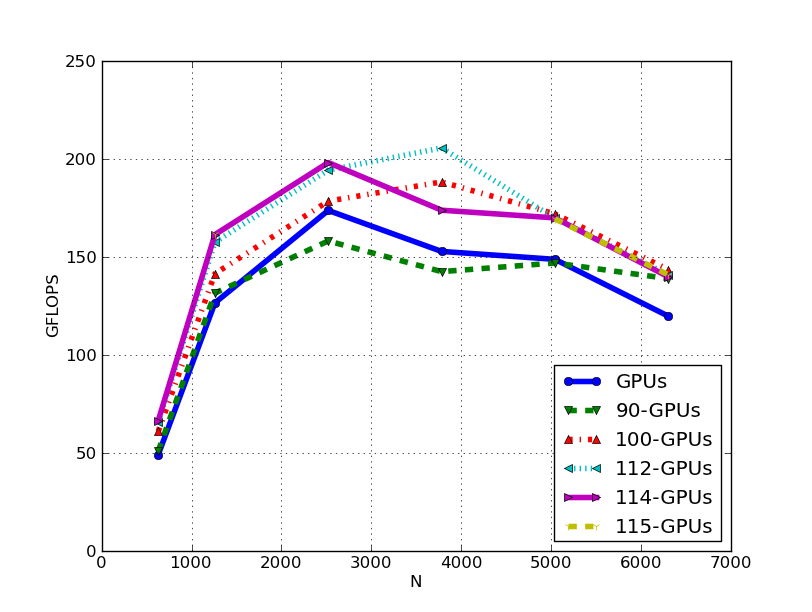}{Parallel GPUs
  performance}{fig:gpus-performance} In Figure
\ref{fig:gpus-performance}, we show the performance when the two GPUs
run concurrently on independent MM on matrices stored continuously in
memory. This performance graphs needs an introduction and explanation:
We took a square problem $N\times N$ and we run it on both GPUs in
parallel. The number of operations are $2*(2N^3)$ and the problem size
can be estimated as $2^{\frac{1}{3}}N \times 2^{\frac{1}{3}}N$. In the
abscissa of the plot we present $2^{\frac{1}{3}}N$.

Now, we notice right the way that the peak performance is about 200
GFLOPS, but instead of increasing as the problem size increases, it
reaches a maximum at about $N=4000$ and then it decreases consistently
and for all configurations. It is like the system reaches a bottle
neck and the throughput get affected negatively by the communication
of data. This makes us believe that, when communications will be
integrand part of the computation as in the following section, the
practical peak performance could be at about 150 GFLOPS. Notice also
that there is no apparent slow down for either one GPU respectively.

In practice, a few configurations are fully stable, and some measures
could not be collected reliably especially for the fastest
configurations such as 115.

\subsection{Accelerators performance}
\label{sec:achievable-performance}

\singlefigure{0.7}{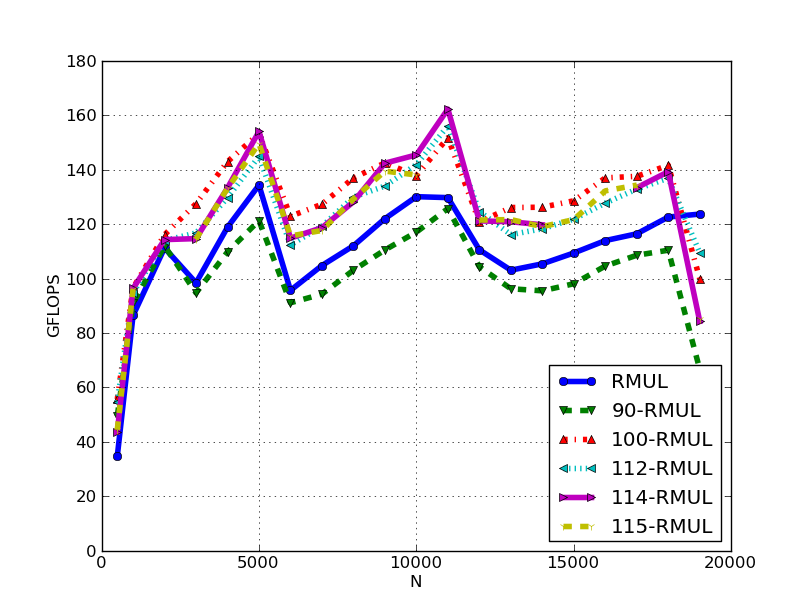}{GPUs Accelerated
  (rmul)}{fig:gpu-accelerated-performance} 

In Figure \ref{fig:gpu-accelerated-performance}, we present the
performance for the recursive algorithm RMUL as we presented in Table
\ref{tab:lmul}.  This figure presents the classic performance curve of
a recursive algorithm: a tooth-saw shape. As a function of the
original problem size, the leaf computation could be
different. Probably, fixed decomposition will have a smaller variance
such as between peaks and valleys. The best performance is about 160
GFLOPS, which is about the peak performance we expected (see previous
Figure \ref{fig:gpus-performance}). Changing the layout of the
operands, when appropriate, could provide smoother performance plots.

\singlefigure{0.7}{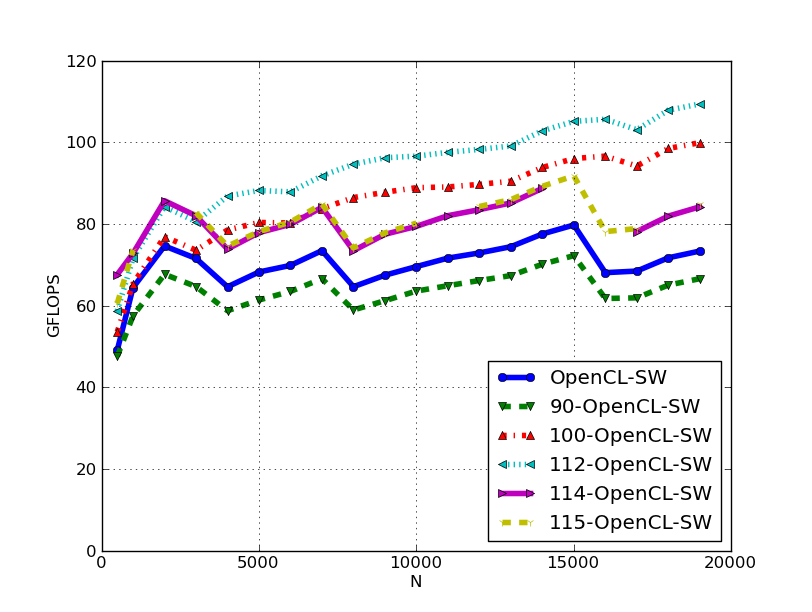}{CPU Winograd Accelerated
  (bmpipe)}{fig:wm-accelerated-performance-0}

Within the OpenCL environment, we measured the performance of the
Winograd's CPU-only MM based on the ATLAS's SGEMM kernel. In Figure
\ref{fig:wm-accelerated-performance-0}, we present the results. We
notice quickly that this picture presents a different performance
plots (more jagged) than what we presented in Figure
\ref{fig:cpus-performance}. At this time, we have no clear explanation
but there could be an interaction between the OpenCL environment and
the GEMM library.

Instead of using the algorithm in Table \ref{tab:rmul}, we could use
the fast recursive algorithm based on the Winograd algorithm. The
advantages of the fast algorithm will be fewer communication and
faster execution time. However, this is beyond the scope of this paper
and we shall address such a optimization in a different work.

\subsection{Conclusions} 

In our system, the APU provides a software solution using only CPUS
that can achieve 90GFLOPS (GotoBLAS). If we would like to improve
performance by just working on a different and fast algorithm, we can
achieve 120 GFLOPS. If we take advantage of both GPUs, we can achieve
sustainable performance of about 160 GFLOPS (and a peak performance of
200 GFLOPS). This is a first attempt in putting together a OpenCL
solution for the implementation of MM using hybrid parallel
systems. The heterogeneous system presents interesting challenges and,
thanks to the OpenCL API, ultimately a flexible and powerful
environment.

\begin{acks}
The authors are in deep dept to the following people who made this
project possible and, most importantly, fun. We thank Matteo Frigo who
made the authors aware about the APU architecture. We thank Chris
Drome for his encouragement. A heartfelt thank goes to Fred Shubert
and the AMD Accelerated Parallel Processing (APP) group who provide an
APU sample and support. We thank also Todd Green for reaching out from
Morgan Kaufmann about OpenCL. Lastly, we thank Matthew Badin,
Alexandru Nicolau, Michael Dillencourt for the conversations about
GPUs.

\end{acks}

\bibliographystyle{acmsmall}
\bibliography{pa.strass2}

%\balancecolumns 

\elecappendix

\medskip
\small

In the following, there are four comments about this work (and reasons
for its rejections EuroPar 2012). We see no need for any rebuttal. 

\begin{verbatim}

======= Review 1 =======

> *** Comments: Comments to author

The method presented in this paper is good. I agree it is suitable for
 that kind of heterogeneous computing environment.

You divide the matrices into four sub-matrices. How better is this in
 case of GPU? This is one of the points we are interested in.

I'm not sure the system works correctly or not with
 over/under-clocking. That can be used to find bottle-neck, however
 not recommended to performance evaluation.

If you are using the advantage of CrossFire, you should mension more
 detail for readers.

Also it is not clear what the CPU threads are doing in two GPU RMUL
 case? Only control and copy/pack/unpack operations?

I could not understand the explanation about the horizontal axis of
 Fig. 3. It seems to exceed the smaller size limit for internal GPU.

The weakness is that, you are using a low-end device for external
 GPU. Usually, we assume external GPU is much faster than internal.

BTW, this is the first time I could not see summary nor conclusion
 section in the paper. Maybe due to lack of pages. We have to estimate
 your main contribution from other parts such as abstract. You wrote
 you can achieved 200GFLOPS in the abstract. But in section 4.2 you
 achieved 200GFLOPS as the summation of independent MM on two
 GPUs. There seems to be some inconsistencies.

======= Review 2 =======

> *** Comments: Comments to author

In this paper the author presents an implementation of a Matrix
 Multiplication for a heterogeneous system. Specifically, the system
 is composed of an Accelerated Processing Unit (APU), which contains a
 processor and a GPU, and an additional GPU.

I am very puzzled with this paper. The author claims that he presents
 a methodology to write code that adapts to different configurations
 of the hardware. With the exception of Table 2, which presents a
 rather intuitive way to decide where to solve a part of the
 computation, I cannot really see any further way in which the
 computation adapts to the hardware configuration. Furthermore, and
 something that I find very important, which part of the code in Table
 2 will execute seems to depend solely on the size of the initial
 matrices. As the matrices are then divided into submatrices, no
 effort is made to decide whether one further step in subdividing
 matrices would lead to a configuration of submatrix sizes that would
 overall provide better performance. This seems to be the case for
 several matrix sizes in Fig. 4.

With respect to the experimental results, although it is mentioned
 that for matrices less than 400x400 the SGEMM of ATLAS or GotoBLAS
 should be used, Fig. 1 presents performance for these libraries for
 matrices larger than 2000x2000.

Overall, I think that this paper does not have a specific target. In
 my opinion it needs a major rewrite in order to reveal this target
 and better explain how it is achieved.

======= Review 3 =======

> *** Comments: Comments to author

I really like the theme of this work, combining multiple GPUs to
 overcome issues in complex applications effectively utilizing the
 larger memory spaces on multiple devices.  The multi-criteria
 optimization is a good target application to motivate this work.

The disappointing part of the paper were the performance results.  I
 found tables 4 and 5 rather disappointing.  First, why is the time on
 the GPU provided in 7 digits of precision, while the CPU and Tcomm is
 only 3 or 4 digits?  This is problematic from an experimental
 methodology.  But besides this, I don't understand the results, and
 there is little explanation for the scaling achieved.  Too much text
 is on the application (neuromophology) and too little on the
 optimization approach and results.

I want to encourage the authors to continue with this work.  Multi-GPU
 work is important and the future for many memory-bound applications
 in HPC.  They an improve on their work with some further analysis of
 the workload.

======= Review 4 =======

> *** Comments: Comments to author

This paper studies the performance of dense matrix-matrix
 multiplication on a system with an APU combining CPU and accelerator
 on die, as well as an external GPU.  While matrix-matrix
 multiplication is only a start for the field, it is definitely a
 valid place to start exploring such systems.

The new area of heterogeneous systems with multiple accelerators of
 varying power and proximity to the host CPU is definitely one worth
 studying.  However, the primary purpose of a paper is to teach
 something to the field.  As I was reading the overall reaction I had
 was, "What is the point?"  The description of the multiplication
 decomposition was written well, but is not new by itself, and there
 was little insight or discussion about how the decomposition
 interacts with the heterogeneous system in new ways.

In general, reading the keys and axis marking of the figures required
 too much strain.  Once the data is understood, I again have to
 question what the relevance of the data is to the field.  What do we
 learn from the figures that expands how we think about matrix
 multiplication, or about heterogeneous systems?

\end{verbatim}

\end{document}

%% file: mydef.tex
\usepackage{times}

\usepackage{pifont}
\usepackage{epsfig}
\usepackage{tabularx}
\usepackage{amssymb}
\usepackage{latexsym}
\usepackage{amsmath}
\usepackage{delarray}
\usepackage{moreverb}
\usepackage{xspace}

\usepackage{lscape}
\usepackage{changebar}
\usepackage{graphicx}
\usepackage{graphics}
\usepackage{mflogo}
\usepackage{xspace}
\usepackage{texnames}
\usepackage{rotating}
\usepackage{alltt}

\usepackage{amsmath}
\usepackage{epsfig}
\usepackage{booktabs,paralist}

\newcommand{\R}{\mathbb{R}}

\newcommand{\Vc}[1]{{\bf #1}}

\newcommand{\Q}[2]{\Vc{#1}_{#2}}

\newcommand{\doublefigure}[5]{{\begin{figure}[ht] %
\centering %
\includegraphics[width=#1\linewidth]{#2}
\includegraphics[width=#1\linewidth]{#3} %
\caption{#4}%
\label{#5}%
\end{figure}}}

\newcommand{\singlefigure}[4]{{\begin{figure}[htp] %
\centering %
\includegraphics[width=#1\linewidth]{#2}\\
\caption{#3}%
\label{#4}%
\end{figure}}}